\documentclass[pra,showpacs,preprint,superscriptaddress,aps]{revtex4-1}
\usepackage[utf8]{inputenc}
\usepackage{amsmath,amssymb,amsfonts}
\usepackage{bm,graphicx,subfigure,float}
\usepackage{longtable,multirow,supertabular,xspace}
\usepackage{color}
\usepackage{epsf,epsfig,graphics}
\usepackage[colorlinks=true, linkcolor=blue, citecolor=red, urlcolor=blue]{hyperref}
\usepackage{orcidlink}
\DeclareUnicodeCharacter{FE20}{\nobreakspace}
\DeclareUnicodeCharacter{FE21}{\nobreakspace}
\begin{document}
\title{Two rotating particles interacting via two-body Gaussian potential and harmonically confined in two spatial dimensions}
\author{Md Hamid 
\orcidlink{0000-0002-2658-5866}
}
\email{hamidjmi2008@gmail.com}
\author{M. A. H. Ahsan 
\orcidlink{0000-0002-9870-2769}
}
\email{mahsan@jmi.ac.in}
\affiliation{Department of Physics, Jamia Millia Islamia (Central University),\
New Delhi 110025, India.}
\begin{abstract}
\indent We investigate a system of two particles harmonically confined in a 2D plane and interacting via two-body Gaussian potential with an externally impressed rotation about an axis perpendicular to the plane of motion. Separating the free motion of the center of mass from the relative motion, we use perturbation theory to analytically obtain the eigenenergy 
in the transcendental form for the relative motion of the pair, in a given relative angular momentum state $l$.
We numerically demonstrate that for the Gaussian interaction potential with interaction strength $g_{2}$ and interaction range $\sigma$, the ground-state energy converges within a subspace of finite number of basis functions. We obtain energies for the ground, the first and the second excited states in the interaction strength regime $-4\leq g_{2}\leq 4$ with various values of  interaction range $\sigma$, for relative angular momenta $l=0$ and $l=1$.
For angular momentum $l=0$ and attractive interaction with $g_{2}<0$, the ground-state energy becomes negative thus forming a bound state. As the interaction strength $g_{2}$ takes further negative values below $-2$, the ground-state energy diverges to $-\infty$ thus forming a tightly-bound pair of particles. 
In contrast, for the angular momentum $l=1$, there is no such divergence of the ground-state energy, instead, 
the ground, the first and the second excited state energies coincide with the corresponding non-interacting values, independent of the value of the interaction strength in the regime $-4\leq g_{2}\leq +4$.
For angular momentum $l=0\mbox{ or } 1$ and repulsive interaction with $g_{2}>0$, the ground-state energy increases with increase in interaction strength as well as interaction range. 
For the ground-state of the pair, the inter-dependence of the average kinetic energy $\langle Ke \rangle$, the average harmonic-potential $\langle V_{ho} \rangle$ and the average interaction energy $\langle V_{int} \rangle$ are analyzed with interaction strength $g_{2}=1$ and interaction range in the regime $0.1\leq \sigma\leq 0.9$, for relative angular momentum $l=0$ and $l=1$.
\end{abstract}
\maketitle
\section{INTRODUCTION \label{intro}}
\indent The ultracold trapped dilute gases\cite{inouye1998observation,bradley1995evidence,anderson1995observation,Busch1998-BUSKAW} of alkali atoms with tunable atom-atom interaction via Feshbach resonance\cite{lange2009determination, salasnich2009solitons,jeszenszki2019smooth} provide a system in the laboratory to experimentally explore a quantum many-body system in a controlled environment. 
The inter-atomic s-wave scattering length takes positive values \cite{inouye1998observation,lange2009determination,salasnich2009solitons,pethick2008bose} for repulsive interaction and negative values\cite{bradley1995evidence} for attractive interaction.
A study of such mesoscopic systems
provides insight into the role of interaction in various exotic properties like nucleation of vortices, superfludity\cite{klaiman2006exact,butts1999predicted}, emergence of quantum chaos\cite{talib2025exact,talib2025spectral,frisch2014quantum}
$etc$, exhibited by the system, and sets the stage for extensions to many-body scenario. 
%
\\
\indent 
An interacting two-particle problem, with harmonic confinement and externally impressed rotation, is analytically tractable which retains the fundamental features of many-body systems. The insight thus gained from the analytical study of such a system leads to an understanding of more intricate quantum many-body systems.\\
\indent The theoretical investigation of two-body problem is relevant to various finite systems in physics like the deuteron problem
\cite{yukawa1935,das2003,HydrogenIsotope,NeutronProton,DeuteronStripping,DeuteronProblem,MACHLEIDT20111,NucDeltaFun},
the pair of valence nucleons outside the inert core of the nucleus, the Cooper's pair problem\cite{Cooper}, the hydrogen atom \cite{dirac1981}, quantum dots\cite{kittel,jeong2001kondo,HAROON} $etc$.
\\
\indent For the dilute trapped atomic gases, the use of the Gaussian interaction potential to model the atom-atom interaction \cite{doganov2013two,Imran_15,Imran_16,Imran_2020,hamid2022vortices,mujal2018fermionic,talib2025exact}, provides a means to study the interplay between the interaction strength and the interaction range and offers an insight into the ground-state characteristics, the excitation spectra and emergence of quantum correlations in the system. The Gaussian interaction potential, unlike the $\delta$-function interaction potential
\cite{Busch1998-BUSKAW,liu2001ground,NucDeltaFun}, 
is smooth and provides control over inter-particle interaction through two parameters, namely, the strength of interaction as measured by the s-wave scattering length and the range of the interaction as measured by the width of the Gaussian. In the limiting case of the interaction range  going to zero, the Gaussian potential reduces to the $\delta$-function potential. \\
\indent In nuclear shell model, 
nucleons are assumed to be confined within the nucleus by an effective mean potential, modelled by a harmonic potential.
In the collective model, the nuclei may have rotational and vibrational motion leading to deformed nuclei, which in extreme cases, may take the shape of an oblate spheroid making it an effective quasi-2D system. \\
\indent In the deuteron problem, the nucleon-nucleon interaction has been modelled by the Gaussian\cite{DeuteronProblem} as well as the $\delta$-function\cite{NucDeltaFun} potentials. The study presented in reference\cite{DeuteronProblem} employs an interaction potential with Gaussian profile which does not reduce to $\delta$-function potential in the limiting case of the width of Gaussian going to zero.
In our present study, we use the form of the Gaussian interaction potential which goes over smoothly to the delta-function potential in the limit of the width of the Gaussian going to zero.\\
\indent In reference\cite{doganov2013two}, the authors present study on harmonically confined system of two particles, interacting via Gaussian potential, in 2D plane with no externally impressed rotation. Our present work extends the above study to a rotating system with quantized (non-zero) angular momenta and varying interaction range.
To capture the features quantitatively, we employ an exact diagonalization approach\cite{ahsan2001rotating} within a truncated Hilbert space, which allows us to extract the low-lying eigenenergy and the corresponding eigenstates as functions of the interaction strength and the interaction range in the Gaussian potential.
\\
\indent The article is organized as follows. In Sec.\ref{TheHamiltonian}, we present our model system for two interacting particles confined in 2D by a harmonic trap. After separating the center of mass motion, we analytically obtain the eigenspectrum for the relative motion of two particles interacting via Gaussian potential, in given relative angular momentum states.
In Sec.\ref{Numerical_Results} we present our numerical results for the internal motion of the pair of particles. In Sec.\ref{Summary} we present a summary of our results. Certain derivations required in Sec.\ref{TheHamiltonian} and Sec.\ref{Numerical_Results} have been presented in the Appendix.\\
\section{The system and the Hamiltonian \label{TheHamiltonian}}
\indent We begin with a system of two particles, each of mass $m$, confined harmonically with $xy$-symmetry to 2D plane, interacting via a two-body Gaussian interaction potential and subjected to an externally impressed rotation about the $z$-axis. The relative motion of the pair is described by the Hamiltonian   
\begin{eqnarray}
{\bf H}_{rel}&=&\underbrace{\frac{{\bf p}^{2}}{2\mu}+\frac{1}{2}\mu{\omega}^{2}{\bf r}^{2}}_{H_{0}}+g_{2}V({r})-{\bf \Omega}\cdot\left({\bf r}\times{\bf p}\right).
\nonumber\\
&=&\underbrace{H_{0}+g_{2}V(r)}_{H}-{\bf \Omega}\cdot\left({\bf r}\times{\bf p}\right).
\label{eq:Hrel}
\end{eqnarray}
as obtained in Eq.(\ref{app:Hrel}) in Appendix \ref{apn:hamil},  where $\omega$ is the frequency for the harmonic confinement in the $xy$-plane and $\Omega$ is the externally impressed rotational velocity about the $z$-axis.
Here $H_{0}$ is the non-interacting hamiltonian in the relative coordinate ${\bf r}\equiv{\bf r}_{1}-{\bf r}_{2}$ with the relative momentum ${\bf p}=\mu\dot{\bf r}$ where $\mu=m/2$ is the reduced mass of the pair, as defined in Appendix \ref{apn:hamil}.
The effective two-body interaction strength in 2D in dimensionless form is given by $g_{2}$, as defined in Eq.(\ref{app:g2f}) in Appendix \ref{ssec:int_pot}. The interaction potential has a Gaussian profile in the relative coordinate
\begin{eqnarray}
V(r)=\left(\frac{1}{\sqrt{2\pi}\sigma}\right)^{2} \ \exp\left[-\frac{{r}^{2}}{2\sigma^{2}}\right],
\label{eq:Vgaus}
\end{eqnarray}
where, $\sigma$ is the interaction range as measured by the width of the Gaussian\cite{Imran_15,Imran_2020,Imran_16,Imran_17,hamid2022vortices}. 
The choice of the Gaussian interaction potential allows us to vary both the interaction strength and the interaction range. The interaction potential used in reference\cite{DeuteronProblem} differs from our choice in Eq.(\ref{eq:Vgaus}) by the premultiplier $\left(\frac{1}{\sqrt{2\pi}\sigma}\right)^{2}$. Further, our interaction potential in Eq.(\ref{eq:Vgaus}) goes over smoothly to the $\delta$-function potential used in \cite{NucDeltaFun} as $\sigma$ goes to zero.
\\ \indent The eigenvalue equation for the non-interacting Hamiltonian $H_{0}$ in Eq.(\ref{eq:Hrel}) is
\begin{eqnarray}
H_{0}u_{n_{r},l}(r\alpha, \phi)=\epsilon_{n_{r},l} u_{n_{r},l}(r\alpha, \phi)\nonumber
\end{eqnarray}
with
\begin{eqnarray}
u_{n_{r},l}(r\alpha,\phi)&=&
\sqrt{\frac{\alpha^{2}n_{r}!}{\pi(n_{r}+|l|)!}}(r\alpha)^{|l|}e^{-\frac{1}{2}r^{2}\alpha^{2}}
e^{il\phi}L^{|l|}_{n_{r}}(r^{2}\alpha^{2})
\\
\label{eq:spwfn}
\mbox{and  } \
    \epsilon_{n_{r},l}&=&
    \left(2n_{r}+|l|+1\right)\hbar\omega 
\\
 n_{r}&=&0,1,2,3,\cdots \ \  \mbox{the radial quantum number} \nonumber\\
l&=&-\infty,\cdots,-1,0,+1,\cdots +\infty \ \mbox{the angular momentum} \nonumber
\end{eqnarray}
where $r$ and $\phi$ are the radial coordinate and the azimuthal angle in $xy$-plane, respectively, while $n_{r}$ and $l$ are the corresponding quantum numbers for the relative motion. The 
$L_{n_{r}}^{|l|}\left(r^{2}\alpha^{2}\right)$ is the Associated Laguerre polynomial with $\alpha\equiv\sqrt{\frac{\mu\omega}{\hbar}}$, the inverse harmonic oscillator length for the relative motion.
\\
\indent To construct the eigensolutions of 
\begin{eqnarray}
	{\textbf{H}}
	\psi_{rel}(r,\phi)=E \psi_{rel}(r,\phi)\label{eq:Hrelpsi}
\end{eqnarray}
for the relative motion of the pair, we expand the variational wavefunction into the linear combination of the single-particle basis states in Eq.(\ref{eq:spwfn}) as $\psi_{rel}(r,\phi)=\sum_{n_r,l} c_{n_{r},l}\ u_{n_{r},l}(r,\phi)$
and determine the variational parameters $\{c_{n_{r},l} \}$ as described in the following. \\
\indent Upon 
projecting Eq.(\ref{eq:Hrelpsi}) onto  the state $u_{n_{r},l}(r\alpha,\phi)$, one obtains
\begin{eqnarray}
&&c_{n_{r},l}(\epsilon_{n_{r},l}-\mbox{E} )+g_{2}\ \sum_{n^{\prime}_{r};l^{\prime}}  \ c_{n^{\prime}_{r},l^{\prime}}\int_{0}^{\infty} \int_{0}^{2\pi}\nonumber \\
&& u^{*}_{n_{r},l}(r\alpha,\phi) V(r) u_{n^{\prime}_{r},l^{\prime}}(r\alpha,\phi) { \ rdr } {d}\phi=0
\label{eq:Sec_eq0}
\end{eqnarray}
where, the matrix element
\begin{eqnarray}
&&I_{n_r;l;n^{\prime}_{r};l^{\prime}}(\sigma)\nonumber \\
&=&\int_{0}^{\infty}\int_{0}^{2\pi} u^{*}_{n_{r},l}(r\alpha,\phi) V(r) u_{n^{\prime}_{r},l^{\prime}}(r\alpha,\phi)\ rdr \ d\phi
\nonumber
\label{eq:Ikk}
\end{eqnarray}
is evaluated analytically in Appendix \ref{apn:IKK}
to obtain in terms of the Gauss-Hypergeometric function ${}_{2}{F}_{1}$ as:
\begin{eqnarray}
I_{n_{r},n^{\prime}_{r};l}(\sigma)&=&\frac{\alpha^{2}}{\pi} \sqrt{\frac{n_{r}!}{(n_{r}+|l|)!}\frac{n^{\prime}_{r}!}{(n^{\prime}_{r}+|l|)!}} \nonumber\\
&&\frac{(2\alpha^{2}\sigma^{2})^{|l|} }{(1+2\alpha^{2}\sigma^{2})^{n_{r}+n^{\prime}_{r} +|l|+1}} \nonumber \\
&& {}_{2}{F}_{1}(-|l|-n^{\prime}_{r},-|l|-n_{r};1,
(2\alpha^{2}\sigma^{2})^{2}).
~~\label{eq:V_intGH}
\end{eqnarray}
For the relative angular momentum $l=0$ with interaction range $\sigma\equiv2^{-1/2}s$ and inverse harmonic oscillator length $\alpha\equiv1/l$, Eq.(\ref{eq:V_intGH}) above reduces to Eq.(7) of the reference\cite{doganov2013two} (with $l$ being the harmonic oscillator length in \cite{doganov2013two}), and is given as
\begin{eqnarray}
I_{n_{r},n^{\prime}_{r};0}(\sigma)\
&=&\frac{\alpha^2}{\pi}\ \frac{1}{(1+2\alpha^{2}\sigma^{2})^{n_{r}+n^{\prime}_{r}+1}}\
\nonumber\\
&& {}_{2}{F}_{1}(-n^{\prime}_{r},-n_{r};1,
(2\alpha^{2}\sigma^{2})^{2})~~~~
\label{eq_V_int_m0}
\end{eqnarray}
In Eq.(\ref{eq:V_intGH}), the Gaussian interaction potential gives rise to off-diagonal matrix-elements between the radial basis functions with quantum numbers $n_{r}$ and $n'_{r}$. 
Equation Eq.(\ref{eq:Sec_eq0}) now reduces to a set of coupled linear equations  in $\{c_{n_{r},l}\}$ given as 
\begin{eqnarray}
c_{n_{r},l}(\epsilon_{n_{r},l}-\mbox{E} )+g_{2} \sum_{n^{\prime}_{r}}c_{n^{\prime}_{r},l} \ I_{n_r,n^{\prime}_{r};l}(\sigma)=0
\label{eq:Sec_eq01}
\end{eqnarray}
\subsection{The eigenenergy for the zero-range potential \label{Contact-potential}}
\indent For the interaction range $\sigma\to 0$, the Gaussian interaction potential in Eq.(\ref{eq:Vgaus}) reduces to the $\delta$-function potential. For the relative angular momentum $l=0$, the matrix element in Eq.(\ref{eq_V_int_m0}) becomes $I_{n_{r},n^{\prime}_{r};0}(0)={\alpha^{2}}/{\pi}$ {\it i.e.}  independent of radial quantum number $n_{r}$. Upon substituting this value of matrix element in Eq.(\ref{eq:Sec_eq01}) with $l=0$, it reduces to
\begin{eqnarray}
c_{n_{r},0}(\epsilon_{n_{r},0}-E) + g_{2}\frac{\alpha^{2}}{\pi}  \underbrace{\sum_{n^{\prime}_{r}} \ c_{n^{\prime}_{r},0}}_{\equiv C}=0.
\label{eq_contt_matt_val}\nonumber
\end{eqnarray}
which further reduces to 
\begin{eqnarray}
c_{n_{r},0}+\frac{g_{2}\alpha^{2}}{\pi}\frac{1}{(\epsilon_{n_{r},0}-E)}C=0.\nonumber
\end{eqnarray}
Summing both sides over $n_{r}$, we obtain for $C\ne0$ 
\begin{eqnarray}
\frac{\pi\hbar\omega_{}}{g_{2}\alpha^{2}}+\sum_{n^{\prime}_{r}}\frac{1}{2n^{\prime}_{r}+1-E/\hbar\omega_{}}=0.
\label{eq_contact_sol}
\end{eqnarray}
The above sum is a harmonic series \cite{boas2006mathematical} and to obtain the eigenenergy of the pair, we truncate the summation to a finite size ${N}_{c}$ of the critical Hilbert space as described in the next section.
\subsection{The eigenenergy with the Gaussian potential
\label{finite_range_potential}}
\indent To obtain the eigenenergy $E$ with the Gaussian interaction potential in Eq.(\ref{eq:Sec_eq0}), the parameter $c_{{n}_{r},l}$ as obtained in Appendix \ref{apn:Cnr} is
\begin{eqnarray}
c_{n_{r},l}=-\frac{g_{2}I_{0,n_{r},l}(\sigma)}{(\epsilon_{n_{r},l}-E)},
\end{eqnarray}
Substituting the above expression for $c_{n_{r},l}$ in Eq.(\ref{eq:Sec_eq01}), one obtains 
\begin{eqnarray}
\frac{\hbar\omega}{g_{2}}I_{n_{r},0;l}(\sigma) &+&
\sum_{n^{\prime}_{r}} \frac{I_{n^{\prime}_{r},0;l}(\sigma)I_{n^{\prime}_{r},n_{r};l}(\sigma)}{\epsilon_{n'_{r},l}/\hbar\omega-E/\hbar\omega}=0.
\label{eq:12}
\end{eqnarray}
For the ground-state eigenenergy, we set $n_{r}=0$
in Eq.(\ref{eq:12}), 
\begin{eqnarray}
\frac{\hbar\omega}{g_{2}}I_{0,0;l}(\sigma) + \sum_{n'_{r}} \frac{I^{2}_{n^{\prime}_{r},0;l}(\sigma)}{\epsilon_{n'_{r},l}/\hbar\omega-E/\hbar\omega}=0,~~~~
\label{eq:I00I0nr}
\end{eqnarray}
where the matrix-element 
in Eq.(\ref{eq:I00I0nr}) as obtained in Appendix \ref{apn:IKK} is
\begin{eqnarray}
I_{n^{\prime}_{r},0;l}(\sigma)&=&\frac{\alpha^{2}}{\pi}\frac{(2\alpha^{2}\sigma^{2})^{|l|}}{(1+2\alpha^{2}\sigma^{2})^{1+n^{\prime}_{r}+|l|}}
\sqrt{\frac{(|l|+n^{\prime}_{r})!}{n^{\prime}_{r}!|l|!}}.\nonumber
\label{eq:I0nrm}
\end{eqnarray}
Using the Lerch transcendental function\cite{erdelyi1953higher} $\Phi\left(z,s,a\right)=\sum_{n}^{}(n+a)^{-s}z^{n}$,
the series $\sum_{n^{\prime}_{r}} \frac{I^{2}_{n^{\prime}_{r},0;l}(\sigma)}{\epsilon_{n^{\prime}_{r},l}/\hbar\omega-E/\hbar\omega}$ in Eq.(\ref{eq:I00I0nr}) can be written as
\begin{eqnarray}
&&\sum_{n^{\prime}_{r}} \frac{I^{2}_{n^{\prime}_{r},0;l}(\sigma)}{\epsilon_{n^{\prime}_{r},l}/\hbar\omega-E/\hbar\omega}\nonumber\\
&=&\frac{\alpha^{4}}{\pi^{2}} \frac{(2\alpha^{2}\sigma^{2})^{2|l|}}{(1+2\alpha^{2}\sigma^{2})^{2(|l|+1)}} \ 
 \left(\frac{(|l|+n^{\prime}_{r})!}{n^{\prime}_{r}!|l|!}\right) \nonumber\\
&& \sum_{n^{\prime}_{r}} \frac{1}{2n^{\prime}_{r}+1+|l|-E/\hbar\omega}
\left(\frac{1}{\left(1+2\alpha^{2}\sigma^{2}\right)^2}\right)^{n^{\prime}_{r}}
\nonumber\\
&=&\frac{\alpha^{4}}{2\pi^{2}}\frac{(2\alpha^{2}\sigma^{2})^{2|l|}}{(1+2\alpha^{2}\sigma^{2})^{2(|l|+1)}} \left(\frac{(|l|+n^{\prime}_{r})!}{n^{\prime}_{r}!|l|!}\right) \nonumber\\
&&  \Phi\left( \frac{1}{\left(1+2\alpha^{2}\sigma^{2}\right)^{2}},1,\frac{1+|l|-\frac{E}{\hbar\omega}}{2}\right).
\end{eqnarray}
The final expression for the ground-state eigenenergy of the pair for the relative motion form Eq.(\ref{eq:12}) is obtained as 
\begin{eqnarray}
\frac{\hbar\omega}{g_{2}}&=&-\frac{\alpha^{2}}{2\pi}\frac{\left(2\alpha^{2}\sigma^{2}\right)^{|l|}}{(1+2\alpha^{2}\sigma^{2})^{(|l|+1)}} 
\left(\frac{(|l|+n^{\prime}_{r})!}{n^{\prime}_{r}!|l|!}\right)\nonumber\\
&& \Phi\left( \frac{1}{\left(1+2\alpha^{2}\sigma^{2}\right)^{2}},1,\frac{1+|l|-\frac{E}{\hbar\omega}}{2}\right).
~~~~\label{eq:final1}
\end{eqnarray}
The expression in Eq.(\ref{eq:final1}) gives the  ground-state eigenenergy and is the main analytical result of our present work.
In this study we take two values of relative angular momentum $l=0$ (even-parity) and $l=+1$ (odd-parity).
Either sign of the relative angular momentum {\it i.e. $l=\pm 1$}, gives the same eigenenergy $E$ {\it i.e.} the system is independent of rotation in either direction in the relative coordinate system.
However, for $l=0$, the Eq.(\ref{eq:final1}) reduces to
\begin{eqnarray}
\frac{\hbar\omega}{g_{2}}
=-\frac{\alpha^{2}}{2 \pi}\frac{1}{(1+2\alpha^{2}\sigma^{2})}
\Phi\left[\frac{1}{(1+2\sigma^2\alpha^2)^{2}},1,\frac{1-E/\hbar\omega}{2}\right], \nonumber\\
\label{eq:final_Lerch}
\end{eqnarray}
which is same as Eq.(17) in reference\cite{doganov2013two}.
\begin{figure}
{\includegraphics[width=1.0\linewidth]{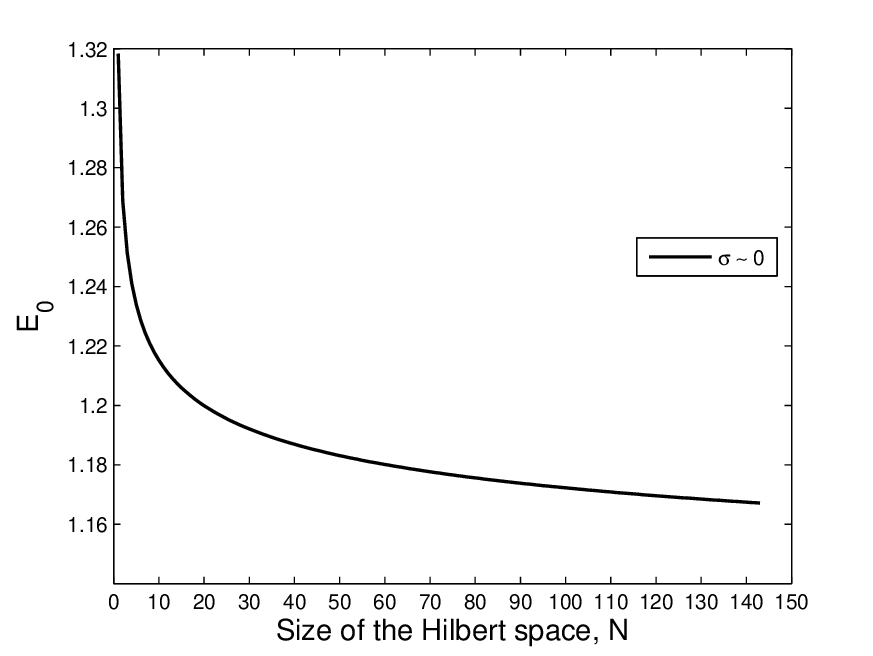}}\caption{\label{Fig:a}(Colour online) Plots for the ground-state energy $E_{0}$ for the relative motion of the pair {\it vs} size of the Hilbert space $N$ for relative angular momentum $l=0$ and interaction strength $g_{2}=+1$ for interaction range $\sigma=0.0001$.}
\end{figure}
\begin{figure}
\includegraphics[width=1.0\textwidth]{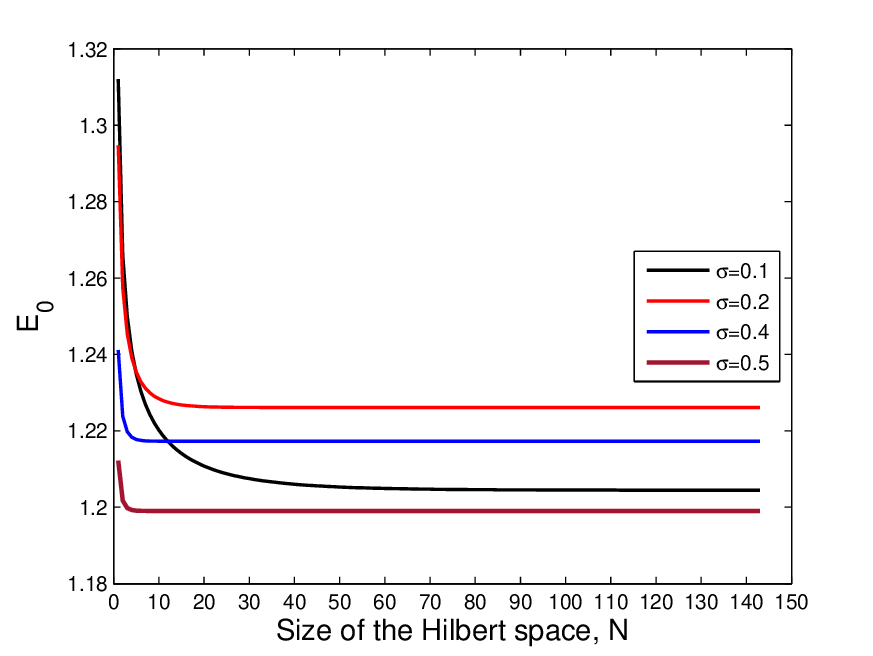}
\caption{\label{EvsNm0}(Colour online) Plots for the ground-state energy $E_{0}$ of the pair {\it vs} the size $N$ of the Hilbert space for relative angular momentum $l=0$, interaction strength $g_{2}=+1$ for various values of interaction range $\sigma=0.1,0.2,0.4,0.5$.}
\end{figure}
\begin{figure}[!htb]
{\includegraphics[width=1.0\linewidth]{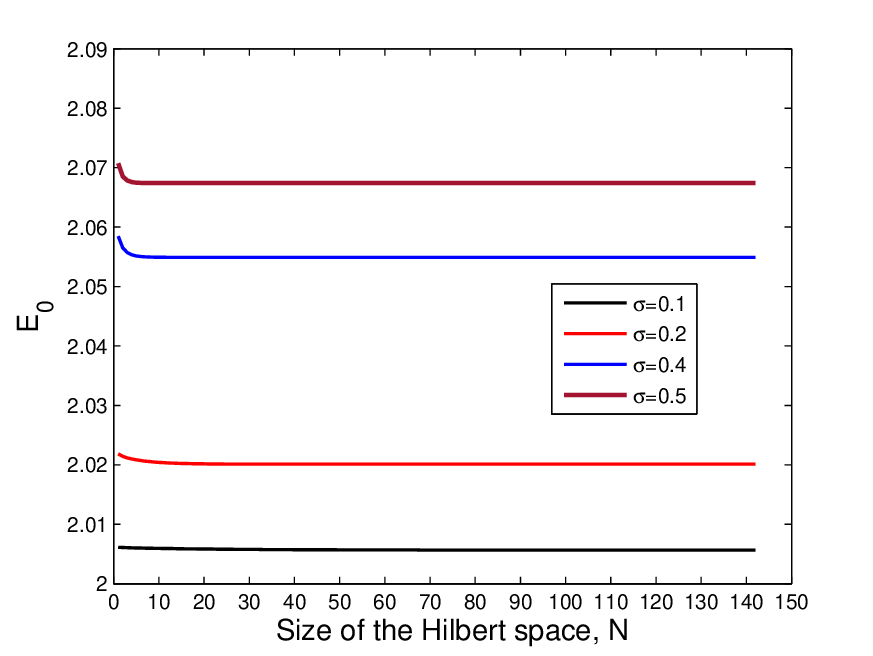}}
\caption{\label{EvsNm1s}(Colour online) The ground-state energy $E_{0}$ {\it vs} size of the Hilbert space $N$, for relative angular momentum $l=1$ with interaction strength $g_{2}=+1$ and various values of interaction range $\sigma=0.1,0.2,0.4,0.5$. It is observed that the size $N_{c}$ of the critical Hilbert space, at which the ground-state energy converges, decreases with increase in interaction range $\sigma$.}
\end{figure}
%
\section{Numerical Results and Discussion
\label{Numerical_Results}}
\begin{figure}[!htb]
{\includegraphics[width=1.0\linewidth]{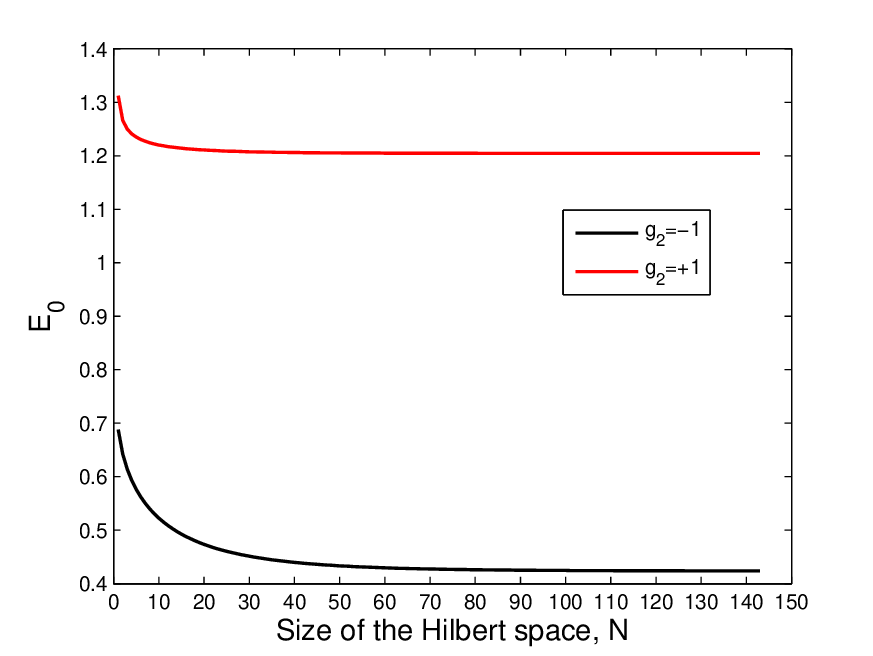}}
\caption{\label{EvsNm1}(Colour online) Plot for the ground-state energy $E_{0}$ of the pair {\it vs} number of basis-states $N$ for both attractive and repulsive interaction with $g_{2}=\mp1$ and interaction range $\sigma=0.1$, for relative angular momentum $l=0$.}
\end{figure}

\begin{figure}[!htb]
{\includegraphics[width=1.0\linewidth]{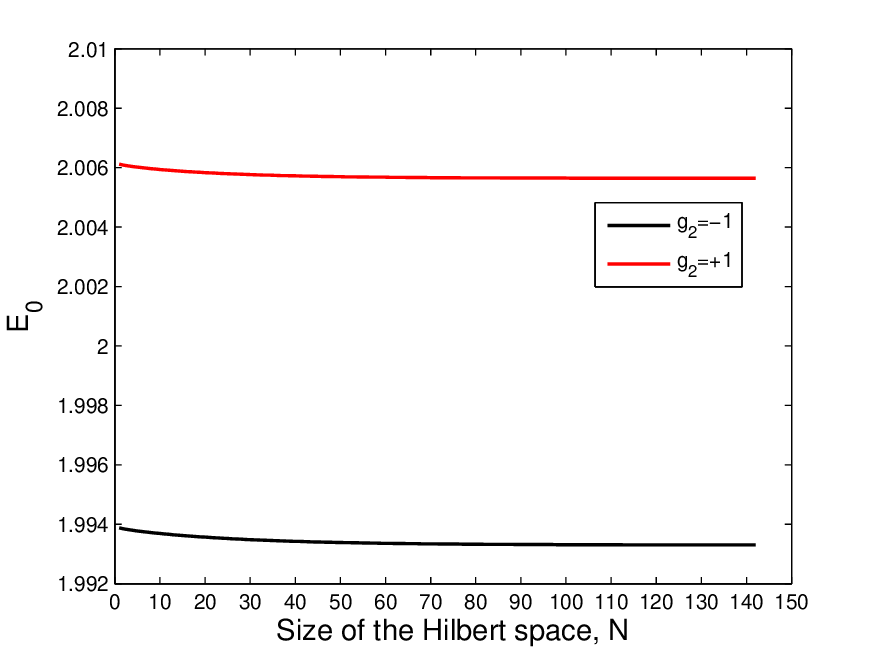}}
\caption{\label{EvsNm1gn1}(Colour online) Plot for the ground-state energy $E_{0}$ {\it vs} number of basis-states $N$ for both attractive and repulsive interaction with $g_{2}=\mp1$ and interaction range $\sigma=0.1$, for relative angular momentum $l=1$.}
\end{figure}

\begin{figure}[!htb]
{\includegraphics[width=1.0\linewidth]{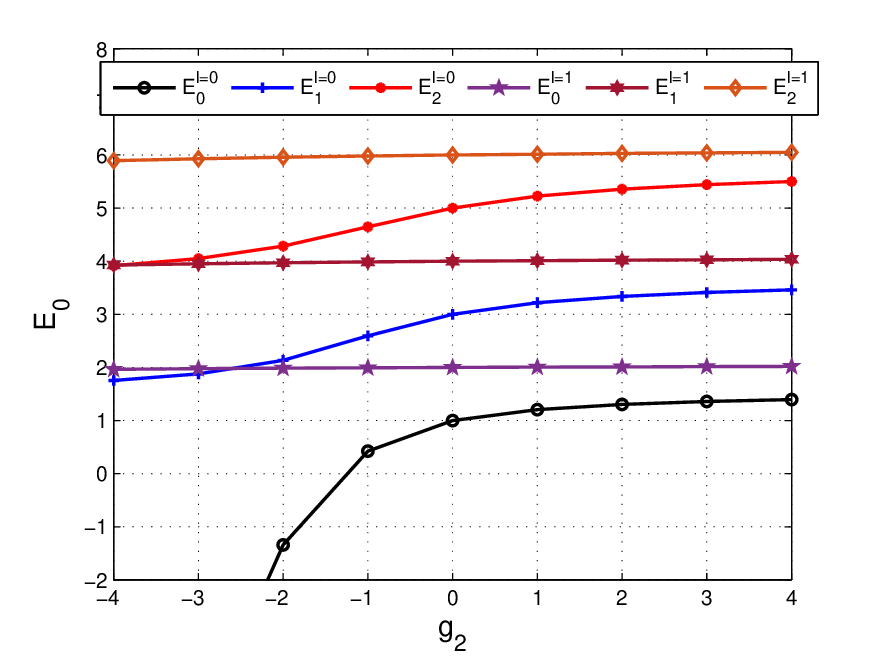}}
\caption{\label{EvsG}(Colour online) Plot for the ground-state, the first and the second excited state energies $\{E_{k}(l),k=0,1,2\}$ for the relative motion of the pair {\it vs} interaction strength in the regime $-4\leq g_{2}\leq 4$ with interaction range $\sigma=0.1$ for angular momentum $l=0,1$.}
\end{figure}

\begin{figure}[!htb]
\includegraphics[width=1.0\linewidth]{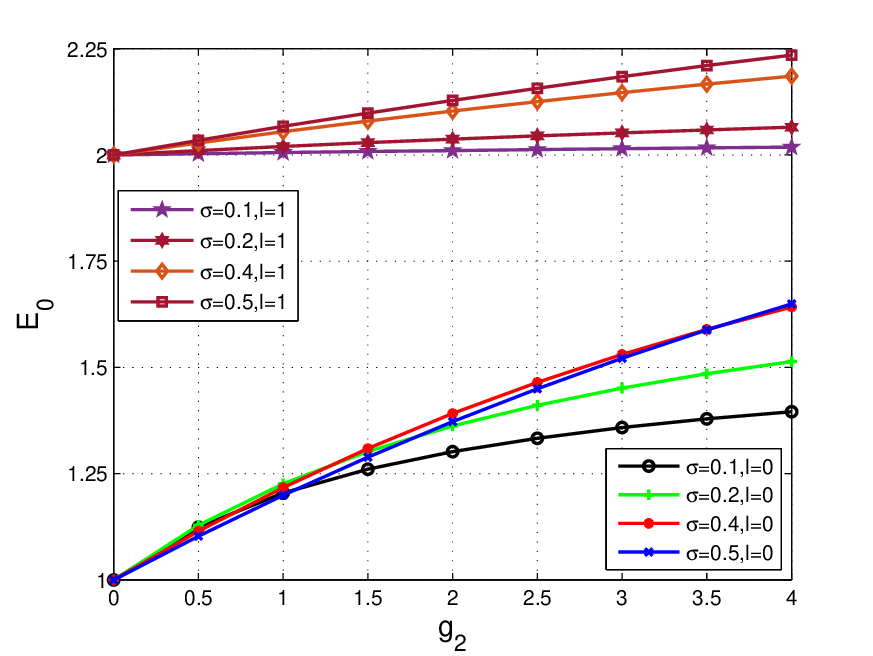}
\caption{\label{figEvsgS}(Color online) Plot for ground-state energy $E_{0}(l,\{g_{2},\sigma\})$ ${\it vs}$ interaction strength $g_{2}$ for various values of interaction range $\sigma$ and angular momentum $l=0,+1$. It is observed that both for $l=0$ and $l=1$, the ground-state energy, in general, increases with increase in interaction strength as well as increase in interaction range. See the text for the discussion.}
\end{figure}

\begin{figure}[!htb]
{\includegraphics[width=1.0\linewidth]{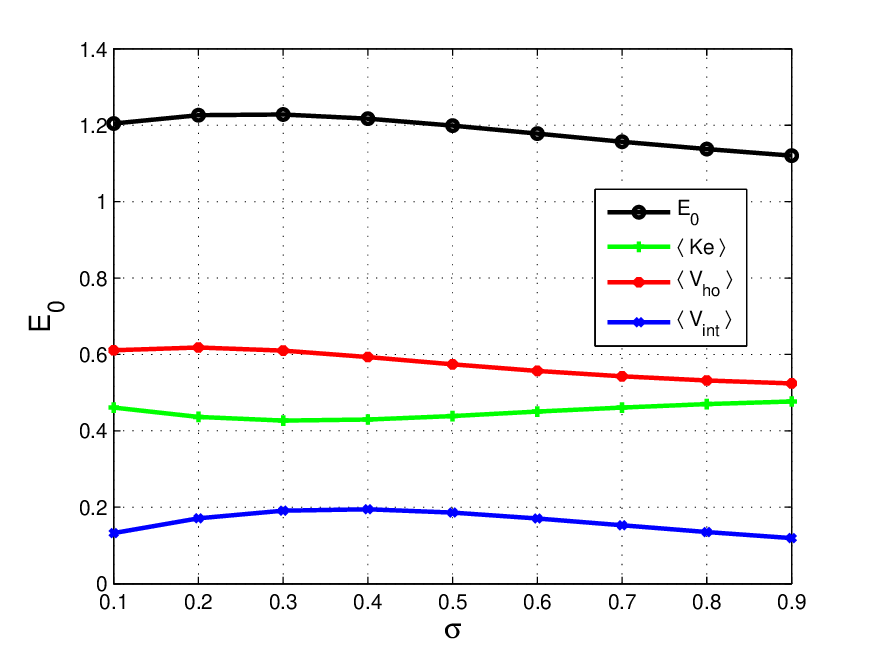}}
\caption{\label{E_V_K_P_vs_SIG_G2_1_M0}(Colour online) Plot for ground-state energy $E_{0}$, average interaction energy $\langle V_{int} \rangle$, average kinetic energy $\langle K_e \rangle$ and average harmonic potential $\langle V_{ho}\rangle$ {\it vs} interaction range $\sigma$ in the regime $0.1\leq\sigma\leq 0.9$, with interaction strength $g_{2}=1$, for angular momentum $l=0$.}
\end{figure}

\begin{figure}[!htb] 
{\includegraphics[width=1.0\linewidth]{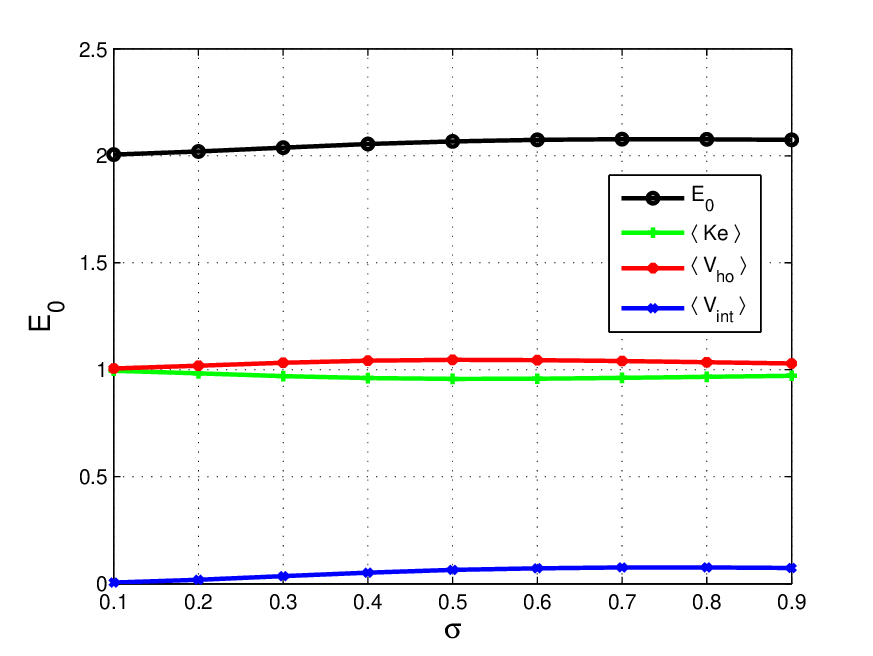}}
\caption{\label{E_V_K_P_vs_SIG_G2_1_M1}(Colour online) Plot for ground-state energy $E_{0}$, average interaction energy $\langle V_{int}\rangle$, average kinetic energy $\langle Ke\rangle$ and harmonic potential $\langle V_{ho}\rangle$ {\it vs} interaction range in the regime $0.1\leq\sigma\leq 0.9$ with interaction strength $g_{2}=+1$ and angular momentum $l=1$.}
\end{figure}

\indent In this section we present our numerical results on the relative motion of a pair of particles, confined harmonically in $xy$-plane and interacting via Gaussian interacting potential with an externally impressed rotation about the $z$-axis. The energy is measured in units of $\hbar\omega$ and the length in units of the harmonic oscillator length $\alpha^{-1}=\sqrt{\frac{\hbar}{\mu\omega}}$, $\alpha$ being the inverse harmonic oscillator length.
The interaction range is varied in the regime $0.1\leq \sigma \leq 0.9$ and the dimensionless effective interaction strength as defined in Eq.(\ref{app:g2f}) in Appendix \ref{ssec:int_pot}, is taken in the regime $-4\leq g_{2}\leq 4$. The ground-state-energy $E_{0}(l)$, for a given value of relative angular momentum $l$, in the non-rotating (laboratory) frame is numerically obtained. The rotational angular velocity for the system to be in angular momentum state $l$ is thus given by $\Omega=\frac{E_{0}(l)-E_{0}(l=0)}{l\hbar}$.
\\
\indent The Hilbert space for the relative motion of a pair of particles is given by the harmonic oscillator basis-functions in Eq.(\ref{eq:spwfn}) as a function of relative coordinates. For a given relative angular momentum $l$, the full Hilbert space for the pair spanned by the basis states $\{u_{n_{r},l}(r,\phi)| n_{r}=0,1,2,\cdots\infty \}$ is infinite-dimensional. 
The subspace spanned by the basis states contributing significantly to the desired eigenstate, for given values of the interaction strength $g_{2}$ and the interaction range $\sigma$ is of finite dimensionality. From now on, we refer to this subspace of the active Fock states\cite{ahsan2001rotating,liu2001ground} 
as the critical Hilbert space and its dimensionality as the size $N_{c}$ of the critical Hilbert space.
\\
\indent For a pair of particles in a given relative angular momentum state $l$, the eigenenergy is given by
\begin{eqnarray}
E(l)&=&\langle \Psi_{rel}(l)|{\bf H}|\Psi_{rel}(l)\rangle=\langle \Psi_{rel}(l)|H_{o}|\Psi_{rel}(l)\rangle
\nonumber\\
&+&g_{2} \underbrace{\langle \Psi_{rel}(l)|V(r)|\Psi_{rel}(l)\rangle}_{{\langle V\rangle}_{\Psi(l)}}
\nonumber \\
&=& \sum_{n_{r}}\underbrace{|c_{n_{r},l}|^{2} \left(2n_{r}+1+|l|\right)}_{\ge 1}+ g_{2}\langle V\rangle _{\Psi(l)}.\nonumber
\end{eqnarray}
With $\sum_{n_{r}}|c_{n_{r},l}|^{2}=1$, the lower bound to the eigenenergy is given by 
\begin{eqnarray}
E(l)\approx \left(2n_{r}+1+|l|\right)+ g_{2}\langle V\rangle _{\Psi_{rel}(l)}.\label{eq:eapp} 
\end{eqnarray}
We will use the eigenenergy given by Eq.(\ref{eq:eapp}) to compare our results with the ones in  reference\cite{doganov2013two}. 
For the pair in the given angular momentum state, we set-up the secular equation and diagonalize the resulting interaction matrix using Davison algorithm to compute a few lowest lying  eigenvalues and the corresponding eigenvectors\cite{ahsan2001rotating}.
\\  
\indent
 We begin with analysing the 
ground-state energy $E_{0}(l)$ in Eq.(\ref{eq_contact_sol}) and Eq.(\ref{eq:final1}) as the number of basis-states $N$ is increased to determine the size $N_{c}$ of the critical Hilbert-space, for given values of the interaction strength $g_{2}$, the interaction range $\sigma$ and the relative angular momentum $l$.
\\  
\indent To examine the convergence of the ground-state energy for the $\delta$-function interaction potential, we numerically take a small value of interaction range $\sigma=0.0001$, which practically corresponds to the $\delta$-function potential. We present the ground-state energy from Eq.(\ref{eq_contact_sol}) {\it vs} the number of basis-states $N$ plotted in Fig.\ref{Fig:a} with interaction strength $g_{2}=+1$ for angualr momentum $l=0$.
We observe that the ground-state energy decreases continuously with increasing number of basis states and does not converge.
In principle, to obtain convergence in the ground-state energy for $\delta$-function potential\cite{PhysRevA.60.1451,Imran_2020}, one requires infinite number of basis states\cite{doganov2013two}.
\\
\indent To examine the dependence of the ground-state energy on the interaction range, with angular momentum $l=0$, we present in Fig.\ref{EvsNm0} the ground-state energy from Eq.(\ref{eq:final1}) {\it vs} the number of basis states $N$ with the interaction strength $g_{2}=1$ and the interaction range
$\sigma=0.1,0.2,0.4,0.5$. We observe that with increase in the interaction range $\sigma$, the convergence in the ground-state energy is achieved for a finite number of basis states $N_{c}$
as summarized in Table.\ref{Tab:EvsS}. 
It is further observed that with increase in interaction range $\sigma$, the ground-state energy converges faster with $N$ $i.e.$ at smaller size $N_{c}$ of the critical Hilbert space.
\\
\indent To further examine the dependence of the ground-state energy on the interaction range, with angular momentum $l=1$, we present in Fig.\ref{EvsNm1s} the ground-state energy {\it vs} number of basis-states $N$, interaction strength $g_{2}=+1$ for various values of interaction range $\sigma=0.1,0.2,0.4,0.5$. 
For angular momentum $l=1$ with negative parity, particles keep apart under the influence of the centrifugal force due to rotation thereby reducing the effect of repulsive interaction.
For interaction range $\sigma=0.1$, the ground-state energy is found to decrease slowly with the number of basis states $N$. 
However, for interaction range $\sigma=0.2$, the ground-state energy decreases with increase in the number of basis states $N$ and converges at $N_{c}=48$. Further, for interaction range $\sigma=0.4,0.5$, the ground-state energy $E_{0}$ decreases with increase in the number of basis states $N$ and converges at $N_{c}=11,8$, respectively. The results are summarized in Table.\ref{Tab:EvsS}.
\\
\begin{table}
	\begin{tabular}{|c|c|c|c|c|}
		\hline
		\multicolumn{5}{|c|}{$g_{2}=+1$} \\ \hline
		$\sigma$&\multicolumn{2}{c|}{$l=0$}&\multicolumn{2}{c|}{$l=1$} \\ \hline
		& $E_{con}$&${N}_{c}$ &$E_{con}$&${N}_{c}$ \\ \hline
		0.1&1.2046  &74  &2.0056 &49  \\
		\hline
		0.2	&1.2261  &25  &2.0200  &48  \\
		\hline
		0.4	& 1.2172 &14  &2.0549  &11  \\
		\hline
		0.5	&1.1990  &12  &2.0673  &8  \\
		\hline
	\end{tabular}
	\caption{\label{Tab:EvsS} For relative angular momentum $l=0,1$, the size $N_{c}$ of the critical Hilbert space for various values of  interaction range $\sigma$ and interaction strength $g_{2}=+1$. It is observed that the size $N_{c}$ of the critical Hilbert space decreases with increase in interaction range $\sigma$. $E_{con}$ is the converged value of the ground-state energy corresponding to the size $N_{c}$ of the critical Hilbert-space.}
\end{table}
\begin{table}[!htb]
\begin{tabular}{|c|cc|cl|} \hline \hline
$g_{2}$&\multicolumn{2}{c|}{$l=0$}&\multicolumn{2}{c|}{$l=1$} \\ \hline
& $E_{con}$&$N_{c}$ &$E_{con}$&$N_{c}$\\
-4 & *    &  * &1.9619  &136 \\
-3 & *    &  * &1.9750  &122  \\
-2 & *    &  * &1.9851  &118  \\
-1 &0.4238 & 139 &1.9933 &46  \\
+1 &1.2046 & 74 &2.0056  &49  \\
+2 &1.3017 & 97 &2.0104  &100  \\
+3 &1.3583 & 97 &2.0146  &139  \\
+4 &1.3957 & 107&2.0184  &88  \\
\hline \hline
\end{tabular}
\caption{The size $N_{c}$ of the critical Hilbert space for interaction strength in the regime $-4\leq g_{2}\leq +4$ and range $\sigma=0.1$ with relative angular momentum $l=0,1$. The saturated ground-state energy $E_{sat}$ increases with increase in the interaction strength $g_{2}$.} 
\label{Tab:Esat_g2}
\end{table}
\indent 
To examine the dependence of the ground-state energy on the interaction strength and the interaction range, we present  
in Fig.\ref{EvsNm1} 
the ground-state energy 
{\it vs} number of basis states $N$
for interaction strength $g_{2}=\pm1$ and interaction range $\sigma=0.1$ for angular momentum $l=0$. 
It is observed that the ground-state energy corresponding to repulsive interaction converges faster with respect to the number of basis-states required for convergence, as compared to the attractive interaction. Similar trend is observed in Fig.\ref{EvsNm1gn1} for relative angular momentum $l=1$. These results are summarized in Table.\ref{Tab:Esat_g2}.
\\
\indent Once the active Fock-space of size (dimensionality) $N_{c}$ spanning the critical Hilbert space is found, the ground-state energy $E_{0}$ obtained in Eq.(\ref{eq:final1}) is analysed for different values of interaction strength $g_{2}$ and interaction range $\sigma$, for given value of angular momentum $l$. In our study presented here, we set the size of the critical Hilbert space $N_{c}=85$.
\\
\indent Fig.\ref{EvsG} presents energies for the ground-state $E_{0}(l)$, the first-excited state ${E_{1}}(l)$ and the second-excited$ E_{2}(l)$, each for angular momentum $l=0,1$ as a function of interaction strength in the regime $-4\leq g_{2}\leq +4$ with $\sigma=0.1$. For attractive interaction $g_{2}<0$ and angular momentum $l=0$ with the eigenstates having positive parities, the ground-state energy becomes negative, thus forming a bound state. 
As $g_{2}$ takes further negative values, the ground-state energy diverges to $-\infty$, as seen in the figure. As the ground-state energy diverges to $-\infty$, the pair of particles turn into a tightly-bound pair which may not obey the same quantum statistics as the constituting particles. 
A possible scenario is two fermions, under strong attractive interaction, form a boson, comprising of tightly-bound pair of fermions.
Further, the first-excited and the second-excited state energies $E_{1}(0)$ and $E_{2}(0)$ decrease with decrease in interaction strength $g_{2}$.
In the repulsive regime with $g_{2}>0$, the energies \{$E_{k}(l=0),k=0,1,2$\} stay above the corresponding non-interacting energies.
\\ 
\indent In contrast, for the rotating case with angular momentum $l=1$, the eigenstates have negative parities $(-1)^{l}$ and the  energies for the ground-state, the first and the second excited  states does not deviate from the corresponding non-interacting energies of 2, 4 and 6 respectively for the interaction strength regime $-4\leq g_{2}\leq 4$. 
\\
\indent To further examine the effect of angular momentum on the ground-state energy, we present in Fig.\ref{figEvsgS} 
the ground-state energy 
$E_{0}(l,\{g_{2},\sigma\})$ {\it vs} interaction strength $g_{2}$ for various values of interaction range $\sigma=0.1,0.2,0.4,0.5$. 
For angular momentum $l=0$, in the interaction strength regime $0\leq g_{2}\leq 1 $, the ground-state energy $E_{0}$ increases linearly starting from $0$ and appears to be independent (curves overlap) of values of interaction range $\sigma$. In the interaction strength regime $1\leq g_{2}\leq 4$, the ground-state energy increases with increase in $g_{2}$ as well as $\sigma$.
For angular momentum $l=1$, whereas for interaction range $\sigma=0.1$, the ground-state energy is independent of the interaction strength $g_{2}$, for the interaction range $\sigma=0.2,0.4,0.5$, the ground-state energy increases with increase in $g_{2}$ as well as $\sigma$.
\\
\indent
We present in
Fig.\ref{E_V_K_P_vs_SIG_G2_1_M0} for angular momentum $l=0$, 
the ground-state energy, the average interaction energy $\langle V_{int}\rangle=g_{2}\langle V(r) \rangle$, the average kinetic energy $\langle Ke\rangle$ and the average harmonic potential $\langle V_{ho} \rangle=\frac{1}{2}\mu\omega^{2} \langle r^{2}\rangle$(calculated in Appendix \ref{app:ho_pot}) 
{\it vs} the interaction range 
$\sigma$ in the regime $0.1\leq\sigma\leq 0.9$ with interaction strength $g_{2}=1$.
The sum of the three average energies equals the ground-state energy $i.e.$ $\langle Ke\rangle+ \langle V_{ho}\rangle+\langle V_{int}\rangle =E_{0}$.
It is observed that, the ground-state energy $E_{0}$ increases with increase in interaction range $\sigma$, attains a maximum at $\sigma=0.2$ and then decreases. Similarly, the average interaction energy $\langle V_{int} \rangle$ increases with increase in interaction range $\sigma$, attains a maximum  at $\sigma=0.4$ and then decreases.
The average kinetic energy $\langle Ke\rangle$ decreases, attains a minimum at $\sigma=0.3$ and then increases with further increase in interaction range $\sigma$.
In contrast, the average harmonic potential $\langle V_{ho} \rangle$ increases with increase in interaction range $\sigma$, attains a maximum at $\sigma=0.2$ and then decreases.
Thus, with increase in interaction range $\sigma$, particles spread out leading to increase in the average harmonic potential $\langle V_{ho} \rangle$ and a corresponding decrease in the average kinetic energy $\langle Ke\rangle$.
\\
\indent We present in Fig.\ref{E_V_K_P_vs_SIG_G2_1_M1} for angular momentum $l=1$, the ground-state energy, the average kinetic energy $\langle Ke\rangle$, the average harmonic potential $\langle V_{ho}\rangle$ and the average interaction energy $\langle V_{int}\rangle$ {\it vs} 
the interaction range
in the regime $0.1\leq\sigma\leq 0.9$ with the interaction strength $g_{2}=1$.
It is observed that, the average interaction energy $\langle V_{int}\rangle$ remains small for small values of interaction range $\sigma$, and attains finite values for $0.3\leq \sigma \leq 0.9$. This is understood as for angular momentum $l=1$,
the centrifugal force due to rotation drives the particles apart, thus reducing the interaction between the particles. 
The average harmonic potential $\langle V_{ho}\rangle$ increases with increase in interaction range $\sigma$, attains a maximum at $\sigma=0.5$ and then decreases with further increase in $\sigma$. 
Correspondingly, the average kinetic energy $\langle Ke\rangle$ decreases with increase in interaction range $\sigma$, attains a minimum at $\sigma=0.5$ and then increases. 
The sum of three average energies equals the ground-state energy $i.e.$ $\langle Ke\rangle+\langle V_{ho}\rangle +\langle V_{int}\rangle=E_{0}$. The ground-state energy is greater than the non-interacting energy $E_{0}(l=1,g_{2}=0)=2$ but stays close to it for all values of $\sigma$ in the regime $0.2\leq \sigma\leq 0.9$. 
\section{Summary and conclusion\label{Summary}}
\indent In summary, we analyzed 
the internal motion of a pair of particles interacting via Gaussian potential and confined harmonically in 2D plane.
A transcendental equation for the eigenenergy of the pair with a given angular momentum was obtained analytically. 
We found numerically that for the $\delta$-function interaction potential, the variational energy for the ground-state of the pair decreases continuously with increasing number of basis functions and does not converge. 
It was found that the use of the Gaussian interaction potential leads to convergence of the ground-state energy within a finite subspace of basis functions spanning the active Fock-space\cite{ahsan2001rotating,liu2001ground} 
for the system.
The energy-spectrum for a given angular momentum was analyzed for various values of interaction strength $g_{2}$ and interaction range $\sigma$.
\\
For angular momentum $l=0$, we observed that the variational energy for the ground-state of the pair decreases faster with increasing number of basis functions for attractive interaction compared to the repulsive interaction
and converges within a finite subspace of basis functions  for a given interaction range. The size of the finite subspace within which the variational ground-state energy converges is smaller for repulsive interaction than for attractive interaction.
For 
attractive interaction $g_{2}<0$, the ground-state energy becomes negative thus forming a bound state whereas
the first and the second excited state energies $E_{1}$ and $E_{2}$ stay positive and increase with increase in interaction strength. 
In the repulsive interaction regime $g_{2}>0$, the energies for the ground, the first and the second excited states,
stay above the corresponding non-interacting energies.
We observed that 
in the interaction strength regime $0\leq g_{2}\leq 1 $, the ground-state energy increases linearly starting from $0$, and appears to be independent of values of interaction range $\sigma$, whereas in the interaction regime $1\leq g_{2}\leq 4$, the ground-state energy increases with increase in $g_{2}$ as well as $\sigma$.
Further, the ground-state energy and the average interaction energy $\langle V_{int} \rangle$ increases with increase in interaction range $\sigma$ and then decreases.
The average kinetic energy $\langle Ke\rangle$ decreases and then increases with increase in interaction range $\sigma$. The average harmonic potential $\langle V_{ho} \rangle$ exhibits a trend complementary to the average kinetic energy $\langle Ke\rangle$.
\\
\indent For angular momentum $l=1$, no bound-state is formed and the energies of the ground, the first and the second excited states coincide with the corresponding non-interacting energies, remaining positive in both the attractive and the repulsive interaction regime.
Further,
for 
interaction range $\sigma=0.1$, the ground-state energy appears independent of interaction strength $g_{2}$ but for interaction range $\sigma=0.2,0.4,0.5$, the ground-state energy increases with increase in interaction strength $g_{2}$ as well as interaction range $\sigma$.
It is also
observed that for 
interaction strength $g_{2}=+1$, the ground-state energy and the average interaction energy $\langle V_{int} \rangle$ increase with increase in interaction range. The average kinetic energy $\langle Ke\rangle$ and the average harmonic potential $\langle V_{ho} \rangle$ exhibit trends similar to $l=0$ case.
\\
\indent In conclusion, our analytical and numerical results on the internal motion of two rotating particles interacting via a tunable Gaussian potential distinctly brings out the effects of the interaction strength, the interaction range and the angular momentum of the pair, on the ground-state and the low-lying excited states of the system. This insight promises to be useful an understanding of more complex quantum many-body systems.
\appendix
\section{The Hamiltonian \label{apn:hamil}}
\indent We consider a system of two particles of mass $m_{1}$ and $m_{2}$ interacting via a two-body potential $V\left(|{r}_{1}-{r}_{2}|\right)\equiv V\left({r}\right)$, confined in $xy$-plane by a symmetric harmonic potential $\left(\omega_{x}=\omega_{y}=\omega\right)$ and subjected to an externally impressed rotation ${\bf \Omega}=\Omega \hat{e}_{z}$ about the $z$-axis. The Hamiltonian for such a system is given by
\begin{eqnarray}
{\bf H}&=&\sum_{i=1}^{2}\left(\frac{{\bf p}^{2}_{i}}{2m_{i}}+\frac{1}{2}m_{i}\omega^{2}{\bf r}^{2}_{i}\right)-{\bf \Omega}\cdot{\sum_{i=1}^{2}{\bf r}_{i}\times{\bf p}_{i}} \nonumber\\
&+&g_{2}\ V(|{\bf r}_{1}-{\bf r}_{2}|),\label{eq:Htotal_app}
\end{eqnarray}
where 
${\bf r}_{i}=x_{i}\hat{e}_{x}+y_{i}\hat{e}_{y}$ and ${\bf p}_{i}=m_{i}(\dot{{x}_{i}}+\dot{{y}_{i}})$ with $ i=1,2$ are the position vectors and the corresponding momenta, respectively, of the particles.
The Hamiltonian can be separated into the center of mass coordinate $
{\bf R}=\frac{m_{1}{\bf r}_{1}+m_{2}{\bf r}_{2}}{m_{1}+m_{2}}$ and the relative coordinate ${\bf r}={\bf r}_{1}-{\bf r}_{2}$ with the total mass $M=m_{1}+m_{2}$ and the reduced mass defined as $\mu=\frac{m_{1}m_{2}}{m_{1}+m_{2}}$.
Further, with the particles momenta ${\bf p}_{1}=m_{1}\dot{\bf R}+\mu\dot{\bf r}$ and ${\bf p}_{2}=m_{2}\dot{\bf R}-\mu\dot{\bf r}$, the relative momentum becomes ${\bf p}=\mu\dot{\bf r}$ and the total momentum 
${\bf P}={\bf p}_{1}+{\bf p}_{2}$.
The Hamiltonian in Eq.(\ref{eq:Htotal_app}) can now be written as
\begin{eqnarray}
{\bf H}&=&\underbrace{\frac{{\bf P}^{2}}{2M}+\frac{1}{2}M{\omega}^{2}{\bf R}^{2}-{\bf \Omega}\cdot\left({\bf R}\times{\bf P}\right)}_{H_{cm}}\nonumber\\ 
&+&\underbrace{\frac{{\bf p}^{2}}{2\mu}+\frac{1}{2}\mu{\omega}^{2}{\bf r}^{2}-{\bf \Omega}\cdot\left({\bf r}\times{\bf p}\right)+g_{2}V(r)}_{H_{rel}},
~~~
\end{eqnarray}
The Hamiltonian is thus separated into the center of mass Hamitonian ${\bf H}_{cm}$ and the relative Hamiltonian ${\bf H}_{rel}$.
\begin{eqnarray}
{\bf H}_{cm}=&\frac{{\bf P}^{2}}{2M}+\frac{1}{2}M{\omega}^{2}{\bf R}^{2}-{\bf \Omega}\cdot{\bf L}
\end{eqnarray}
The Hamiltonian ${\bf H}_{cm}$ describes the free harmonic oscillator in $2D$ with externally impressed rotation about the $z$-axis whose solution $H_{cm} U_{n_{R},L}(R\alpha_{cm},\Phi)=E U_{n_{R},L}(R\alpha_{cm},\Phi)$ is known to be
\begin{eqnarray}
U_{n_{R},L}(R\alpha_{cm},\Phi)&=&
\sqrt{\frac{\alpha_{cm}^{2}n_{R}!}{\pi(n_{R}+|L|)!}}(R\alpha_{cm})^{|L|}e^{-\frac{1}{2}(R\alpha_{cm})^{2}}
\nonumber\\&\times&
e^{i L\Phi} L^{|L|}_{n_{R}}(R\alpha_{cm})^{2}.
\label{eq:spwfnC.M}
\end{eqnarray}
where $n_{R}$ and $L$ are the principal quantum number and the total angular momentum quantum number, respectively, and $L^{|L|}_{n_{R}}(R\alpha_{cm})^{2}$ is the Associated Laguerre polynomials in center of mass coordinates with inverse harmonic oscillator length $\alpha_{cm}=\sqrt{\frac{M\omega}{\hbar}}$ for the center of mass.
\\
\indent The Hamiltonian 
\begin{eqnarray}
{\bf H}_{rel}=\overbrace{\underbrace{                                                                                                                                                                                                                                                                                                                                                                                                                                                                                                                                                                                                                                   \frac{{\bf p}^{2}}{2\mu}+\frac{1}{2}\mu{\omega}^{2}{\bf r}^{2}}_{H_{0}}+g_{2}V({r})}^{H}-{\bf \Omega}\cdot\left({\bf r}\times{\bf p}\right).
\label{app:Hrel}
\end{eqnarray}
describes the relative motion of a pair of two particles discussed in the main text.
The total solution for the Hamiltonian in Eq.(\ref{eq:Htotal_app}) is given by $\Psi_{total}=U_{n_{R},L}(R\alpha_{cm},\Phi)\psi_{rel}$ for the pair.
\section{\label{ssec:int_pot} The effective Gaussian interaction potential in two dimensions}
\indent The Gaussian interaction potential in $3D$,
 with $xy$-symmetry({$\sigma_{x}=\sigma_{y}=\sigma_{\perp}$}),
where $\sigma_{\perp}$($\sigma_{z}$) is two-particle interaction range in $xy$-plane(along $z$-axis), and $r_{\perp}\equiv\sqrt{x^{2}+y^{2}}$, is given by
\begin{eqnarray}
U_0(\mathbf{r-r^{\prime}})&=&
^{3D}{g_{2}}\left(\frac{1}{\sqrt{2\pi}\sigma_{\perp}}\right)^{2} \exp\left[ -\frac{(r_{\perp}-r_{\perp}^{\prime})^{2}}{2\sigma^{2}_{\perp}} \right] \nonumber\\
&& \times \left(\frac{1}{\sqrt{2\pi}\sigma_{z}}\right) \exp \left[-\frac{\left(z-z^{\prime}\right)^{2}}{2\sigma^{2}_{z}} \right] \nonumber
\label{eq:int_pot}
\end{eqnarray}
where $^{3D}{g_{2}}=\frac{4\pi\hbar^{2} a_{s}}{M}$ is the interaction strength with $a_{s}$ the s-wave scattering length. 
With $\lambda_{z}=\frac{\omega_{z}}{\omega_{\perp}}\gg 1$, where $\omega_{\perp}$ and $\omega_{z}$ are the radial and the axial trap frequencies, respectively and particles are confined to quasi-$2D$ plane with no excitation along the $z$-axis.  
Tracing out the $z$-degree of freedom, setting $n_{z}=0$, the effective potential in  $2D$ is obtained as 
\begin{eqnarray}
&&\int {\bf u}_{(n_{z}=0)}(z){\bf u}_{(n_{z}=0)}(z^{\prime}) U_0({\bf r-r^{\prime}}){\bf u}_{(n_{z}=0)}(z)\nonumber\\
&\times&{\bf u}_{(n_{z}=0)}(z^{\prime})dzdz^{\prime}= {^{3D}g_{2}}\frac{1}{\sqrt{2\pi}}\sqrt{\frac{1}{a_{z}^{2}(1+(\sigma_{z}/a_{z})^{2})}}
\nonumber \\
&\times& \left(\frac{1}{\sqrt{2\pi}\sigma_{\perp}}\right)^{2}\exp({-\frac{1}{2\sigma^{2}_{\perp}}}{(r_{\bot}-r_{\bot}^{\prime})^{2}
})\nonumber
\label{app:eqRedG}
\end{eqnarray}
where the harmonic oscillator length along $z$-axis for the confining potential $a_{z}=\sqrt{\frac{\hbar}{m\omega_{z}}}=\sqrt{\frac{\hbar}{m\omega_{\perp}\lambda_{z}}}=\sqrt{\frac{a_{\perp}}{\lambda_{z}}}$
with $\omega_{z}=\omega_{\perp}\lambda_{z}$ leads to   
\begin{eqnarray}
U_{\mbox{effective}}(|{r_{\perp}-r_{\perp}^{\prime}}|)
&=&{^{3D}g_{2}}\frac{1}{\sqrt{2\pi}}\sqrt{\frac{\lambda_{z}}{a^{2}_{\perp}}\frac{1}{(1+\lambda_{z}(\frac{\sigma_{z}}{a_{\perp}})^{2})}} \nonumber\\ &&\left(\frac{1}{\sqrt{2\pi}\sigma_{\perp}}\right)^{2}\exp({-\frac{1}{2\sigma^{2}_{\perp}}}{(r_{\bot}-r_{\bot}^{\prime})^{2}
}).~~~~~~~~~\nonumber
\label{eq:2}
\end{eqnarray}
For $\frac{\sigma_{z}}{a_{\perp}}\ll1$, the dimensionless effective potential in $2D$ becomes 
\begin{eqnarray}
U_{\mbox{effective}}(|{r_{\perp}-r_{\perp}^{\prime}}|)&=&4\pi\frac{a_{s}}{a_{\perp}}\sqrt{\frac{\lambda_{z}}{2\pi}}
\left(\frac{1}{\sqrt{2\pi}\sigma_{\perp}}\right)^{2} \nonumber\\
&&
\exp({-\frac{1}{2\sigma^{2}_{\perp}}}{(r_{\bot}-r_{\bot}^{\prime})^{2}
})\nonumber\\
&=&g_{2} \ \underbrace{\left(\frac{1}{\sqrt{2\pi}\sigma_{\perp}}\right)^{2}\exp({-\frac{1}{2\sigma^{2}_{\perp}}}{(r_{\bot}-r_{\bot}^{\prime})^{2}
})}_{V(r)}
\nonumber
\label{eq:3}
\end{eqnarray}
where  the dimensionless effective two-body interaction strength in 2D is given by
\begin{eqnarray}
g_{2}\equiv4\pi\frac{a_{s}}{a_{\perp}}\sqrt{\frac{\lambda_{z}}{2\pi}} \label{app:g2f}.
\end{eqnarray}
The interaction potential has a Gaussian profile $V(r)$ in the relative coordinate $r\equiv|{r}_{\perp}-{r}_{\perp}^{\prime}|$ with the range of the interaction $\sigma_{\perp}$ in the Gaussian. 
Thus the effective inter-particle interaction in 2D is given by $g_{2}V(r)$.

\section{Calculation of interaction matrix elements \label{apn:IKK} }
\indent In this appendix we outline the steps for calculation of the matrix elements used in the 
secular equation in the main text in Eq.(\ref{eq:Sec_eq0}) for that we consider a normalized single particle wavefunction $u_{n_{r},l}(r\alpha,\phi)$ given in Eq.(\ref{eq:spwfn}), in relative coordinates and integrate it over $rdrd\phi$, where, $r$ is spatial relative co-ordinate and $\phi$ is an azimuthal angle. The single particle interaction elements are being calculated in the following manner:
\begin{eqnarray}
I_{n^{\prime}_{r},l^{\prime};n,l}(\sigma)=\langle u_{n^{\prime}_{r},l^{\prime}}(r_{}\alpha_{},\phi)|V(r)|u_{n_{r},l}(r_{}\alpha_{},\phi) \rangle \nonumber
\end{eqnarray}
with substituting the single particle wavefunction $u_{n_{r},l}(r_{}\alpha_{},\phi)$ 
into the above equation,
\begin{eqnarray}
&&I_{n^{\prime}_{r},l^{\prime};n,l}(\sigma)=\int_{0}^{\infty}\!\!\int_{0}^{2\pi} r_{}\,\mbox{d}r_{}\,\mbox{d}\phi \nonumber\\ 
&\times&\sqrt{\frac{\alpha^{2}_{}n_{r}!}{\pi(n_{r}+|l|)!}} (r_{}\alpha_{})^{|l|}e^{-1/2r_{}^{2}\alpha_{}^{2}}e^{il\phi}\nonumber\\
&\times&L^{|l|}_{n_{r}}(r_{}^{2}\alpha_{}^{2}) \ e^{-\frac{r^{2}}{2\sigma^{2}}} \times
\frac{1}{2\pi\sigma^{2}} \sqrt{\frac{\alpha^{2}_{}n'_{r}!}{\pi(n'_{r}+|l'|)!}} \nonumber\\ 
&\times&(r_{}\alpha_{})^{|l'|}e^{-1/2r_{}^{2}\alpha_{}^{2}}e^{-il'\phi}L^{|l'|}_{n'_{r}}(r_{}^{2}\alpha_{}^{2})
\end{eqnarray}
the integration is being carried out for $\phi$ which gives $\delta$-function,
\begin{eqnarray}
&&I_{n^{\prime}_{r},l^{\prime};n,l}(\sigma)=\frac{1}{2\pi\sigma^{2}}\ \int\!\! r_{}\,\mbox{d}r_{}\,2\pi\delta(l-l') e^{-\frac{r^{2}}{2\sigma^{2}}}
\nonumber\\
&\times&
\sqrt{\frac{\alpha^{2}_{}n_{r}!}{\pi(n_{r}+|l|)!}}
(r_{}\alpha_{})^{|l|}e^{-1/2r_{}^{2}\alpha_{}^{2}}L^{|l|}_{n_{r}}(r_{}^{2}\alpha_{}^{2})\nonumber\\ 
&\times&\sqrt{\frac{\alpha^{2}_{}n'_{r}!}{\pi(n'_{r}+|l'|)!}} \
(r_{}\alpha_{})^{|l'|}e^{-1/2r_{}^{2}\alpha_{}^{2}}L^{|l'|}_{n'_{r}}(r_{}^{2}\alpha_{}^{2})~~~~
\end{eqnarray}
we substitute $l=l'$ and 
gets the following results
\begin{eqnarray}
&&I_{n^{\prime}_{r},n;l}(\sigma)=
\sqrt{\frac{n_{r}!}{\pi(n_{r}+|l|)!}} \sqrt{\frac{n'_{r}!}{\pi(n'_{r}+|l|)!}} \ \nonumber\\
&\times& \pi \frac{1}{2\pi\sigma^{2}} \int\ \underbrace{\alpha^{2}_{}2r_{}\,\mbox{d}r_{}}_{\mbox{d}(r^{2}_{}\alpha^{2}_{})}
(r_{}\alpha_{})^{2|l|}\ \exp{(-r^{2}_{}\alpha^{2}_{}(1+\frac{1}{2\alpha^{2}\sigma^{2}}))} \nonumber\\
&\times& L^{|l|}_{n'_{r}}(r_{}^{2}\alpha_{}^{2}) L^{|l|}_{n_{r}}(r_{}^{2}\alpha_{}^{2})\nonumber
\end{eqnarray}
let $1+\frac{1}{2\alpha^{2}\sigma^{2}}=\Lambda$, another dimensionless quantity
\begin{eqnarray}
&&I_{n^{\prime}_{r},n_{r};l}(\sigma)=
\sqrt{\frac{n_{r}!}{\pi(n_{r}+|l|)!}} \sqrt{\frac{n'_{r}!}{\pi(n'_{r}+|l|)!}} \ \pi \frac{1}{2\pi\sigma^{2}}\nonumber\\
&& \int\ \mbox{d}(\alpha^{2}_{}r^{2}_{} )
(r_{}\alpha_{})^{2|l|}
\exp{(-r^{2}_{}\alpha^{2}_{}\Lambda)}
L^{|l|}_{n'_{r}}(r_{}^{2}\alpha_{}^{2})L^{|l|}_{n_{r}}(r_{}^{2}\alpha_{}^{2})~~~~~~~~\nonumber
\end{eqnarray}
again, we define $r^{2}_{}\alpha^{2}_{}\Lambda =\rho$ in the form of a new dimensionless parameter and substituting it in the above expression
which further simplifies to
\begin{eqnarray}
&&I_{n^{\prime}_{r},n_{r};l}(\sigma)=
\frac{1}{2\pi\sigma^{2}\Lambda}\ \sqrt{\frac{n_{r}!}{(n_{r}+|l|)!}\frac{n'_{r}!}{(n'_{r}+|l|)!}} \nonumber \\
&\times& \int\ \mbox{d}\rho ( \frac{\rho}{\Lambda})^{|l|}
\exp{(-\rho)}
L^{|l|}_{n'_{r}}(\frac{\rho}{\Lambda})L^{|l|}_{n_{r}}(\frac{\rho}{\Lambda})
\label{app:eqmulti}
\end{eqnarray}
using  multiplication formula for the Associated Laguerre polynomials \cite{tcheutia2016coefficients}
\begin{eqnarray}
L_{n_{r}}^{|l|} (\Lambda x)=\sum_{k=0}^{n_{r}} \frac{\Lambda^{k}(|l|+1)_{n_{r}}(1-\Lambda)^{n_{r}-k}}{(|l|+1)_{k}(n_{r}-k)!} L^{|l|}_{k}(x) \nonumber 
\end{eqnarray}
in Eq.(\ref{app:eqmulti}) to make it solvable by removing $1/\Lambda$ term from it. Rewriting Eq.(\ref{app:eqmulti}) in terms of Pochhammer symbols
$(a)_{n}$\cite{arfken2011mathematical}
\begin{eqnarray}
&&I_{n^{\prime}_{r},n_{r};l}(\sigma)=\frac{1}{2\pi\sigma^{2}\Lambda^{1+|l|}}\ \sqrt{\frac{n_{r}!}{(n_{r}+|l|)!}\frac{n'_{r}!}{(n'_{r}+|l|)!}}\nonumber \\
&\times&\sum_{k,k'=0}^{\mbox{min}({n}_{r},{n}^{\prime}_{r})} \frac{\Lambda^{-k-k'}(|l|+1)_{n_{r}}(1-\Lambda^{-1})^{n_{r}+n^{\prime}_{r}-k-k'}}{(|l|+1)_{k}(n_{r}-k)!}\nonumber \\
&\times&\frac{(|l|+1)_{n^{\prime}_{r}}}{(|l|+1)_{k'}(n^{\prime}_{r}-k')!}
\ \int_{0}^{\infty} \mbox{d}\rho \ {\rho}^{|l|}\exp{(-\rho)} L^{|l|}_{k}(\rho)L^{|l|}_{k'}(\rho)\nonumber
\end{eqnarray}
using the orthogonality condition for the Associated Laguerre polynomials\cite{arfken2011mathematical},
%
%
the integrand becomes,
\begin{eqnarray}
&&I_{n^{\prime}_{r},n_{r};l}(\sigma)=\frac{1}{{2\pi\sigma^{2}}\Lambda^{1+|l|}}\ \sqrt{\frac{n_{r}!}{(n_{r}+|l|)!}}\nonumber\\
&\times&\sqrt{\frac{n'_{r}!}{(n'_{r}+|l|)!}} \ \
(\frac{\Lambda-1}{\Lambda})^{n_{r}+n^{\prime}_{r}} \nonumber\\
&\times& \sum_{k=0}^{\mbox{min} ({n}_{r},{n}^{\prime}_{r})} \frac{(|l|+1)_{n_{r}}}{(|l|+1)_{k}(n_{r}-k)!}\nonumber \\
&\times&\frac{(|l|+1)_{n^{\prime}_{r}}}{(|l|+1)_{i}(n^{\prime}_{r}-k)!}\frac{(k+|l|)!}{k!}(\frac{1}{(\Lambda-1)^2})^{k}
\end{eqnarray}
again, writing the Pochhammer symbol  
$(a)_{n}=\frac{(a+n-1)!}{(a-1)!}$ in factorials
\begin{eqnarray}
&&I_{n^{\prime}_{r},n_{r};l}(\sigma)=\frac{1}{{2\pi\sigma^{2}}\Lambda^{1+|l|}}\ \sqrt{\frac{n_{r}!}{(n_{r}+|l|)!}} \nonumber \\ 
&\times& \sqrt{\frac{n'_{r}!}{(n'_{r}+|l|)!}} \left(\frac{\Lambda-1}{\Lambda}\right)^{n_{r}+n^{\prime}_{r}} \nonumber\\
&\times& \sum_{k=0}^{\mbox{min}({n}_{r},{n}^{\prime}_{r})}
		\left(\begin{array}{lr}
		|l|+n_{r}  \\
		|l|+k
	\end{array}     \right)
 \left(\begin{array}{lr}
		|l|+n^{\prime}_{r}  \\
		|l|+k
	\end{array}  \right)
  \nonumber\\
 &\times& \ \frac{(|l|+k)!}{k!}\ \left( \frac{1}{(\Lambda-1)^2}\right)^{k}
\label{apn:Inrnrpm}
\end{eqnarray}
and substituting the value of $\Lambda$ in Eq.(\ref{apn:Inrnrpm}) to arrive at the final result
\begin{eqnarray}
I_{n^{\prime}_{r},n_{r};l}(\sigma)
&=&\frac{\alpha^{2}}{\pi} \frac{(2\alpha^{2}\sigma^{2})^{|l|} }{(1+2\alpha^{2}\sigma^{2})^{n_{r}+n^{\prime}_{r} +|l|+1}}\nonumber\\
&& \sqrt{\frac{n_{r}!}{(n_{r}+|l|)!}\frac{n^{\prime}_{r}!}{(n^{\prime}_{r}+|l|)!}}\nonumber \\
&& \sum_{k=0}^{\mbox{min}(n_{r},n^{\prime}_{r})}
\left(\begin{array}{lr}
|l|+n_{r}\nonumber \\
|l|+k
\end{array} \right)
\left(\begin{array}{lr}
|l|+n^{\prime}_{r}\nonumber \\
|l|+k
\end{array} \right)\nonumber\\
&&
\frac{(|l|+k)!}{k!}\ (2\alpha^{2}\sigma^{2})^{2k}.
\label{eq:V_int}
\end{eqnarray}
\section{Calculation of $c_{n_{r},l}$ \label{apn:Cnr}}
In this section, we obtain the parameter
$c_{n_{r},l}$ in the variational wavefunction $|\psi_{rel}\rangle=\sum c_{n_{r},l}|u_{n_{r},l}\rangle$ using perturbation theory. 
We begin with the eigenvalue equation $H_{rel}|\psi_{rel}\rangle=E|\psi_{rel}\rangle$ and using Eq.(\ref{eq:Hrel}), we obtain
\begin{eqnarray}
&&({H}_{rel}-H_{0})|\psi_{rel}\rangle = g_{2}V(r)|\psi_{rel}\rangle, \nonumber.
\end{eqnarray}
Upon projecting the above equation onto the basis-state $|u_{n_{r},l}\rangle$, we get:
\begin{eqnarray}
({E}-\epsilon_{n_{r},l})\underbrace{\langle u_{n_{r},l}|\psi_{rel}\rangle}_{\equiv c_{n_{r},l}} &=&g_{2} \langle u_{n_{r},l}|V(r)|\psi_{rel}\rangle\nonumber\\
\Rightarrow \ c_{n_{r},l} &=& 
\frac{g_{2} \langle u_{n_{r},l}|V(r)|\psi_{rel}\rangle}{E-\epsilon_{n_{r},l}}.~~~~~
\label{eq:apncnr}
\end{eqnarray}
Till now, no approximation has been made.
To obtain the ground-state in the first approximation in a given angular momentum state $l$, we set $|\psi_{rel}\rangle =|u_{{0},l}\rangle$ in Eq.(\ref{eq:apncnr}),
\begin{eqnarray}
c_{n_{r},l}=-\frac{g_{2}\overbrace{ \langle u_{n_{r},l}|V(r)|u_{{0},l}\rangle}^{I_{0,n_{r};l}(\sigma)}  }{(\epsilon_{n_{r},l}-E)}=\frac{-g_{2}I_{0,n_{r};l}(\sigma)}{\epsilon_{n_{r},l}-E}.
\label{app:cnrol}
\end{eqnarray}
For the excited-states, we set $|\psi_{rel}\rangle=|u_{n'_{r},l}\rangle$ to obtain
\begin{eqnarray}
c_{n_{r},l} = -\frac{g_{2} \langle u_{n_{r},l}|V(r)|u_{n'
		_{r},l}\rangle}{(\epsilon_{n_{r},l}-E)}.
\end{eqnarray}
%
\section{Expectation of the trap potential.
\label{app:ho_pot}}
The expectation value of the symmetric harmonic trap potential is given by
\begin{eqnarray}
\langle \alpha^{2}r^{2}\rangle =&&\int_{0}^{\infty}\!\!\int_{0}^{2\pi}  r_{}\,\mbox{d}r_{}\,\mbox{d}\phi \sqrt{\frac{\alpha^{2}_{}n_{r}!}{\pi(n_{r}+|l|)!}} \nonumber\\ 
&\times&(r_{}\alpha_{})^{|l|}e^{-1/2r_{}^{2}\alpha_{}^{2}}e^{il\phi}\nonumber\\
&\times&L^{|l|}_{n_{r}}(r_{}^{2}\alpha_{}^{2}) \times
(r^{2}\alpha^{2}) \sqrt{\frac{\alpha^{2}_{}n'_{r}!}{\pi(n'_{r}+|l'|)!}}\nonumber \\ 
&\times&\ \ (r_{}\alpha_{})^{|l'|} e^{-1/2r_{}^{2}\alpha_{}^{2}}
e^{-il'\phi}.\nonumber
\end{eqnarray}
The integration over the angle $\phi$ gives a factor of $2\pi \delta_{l,l'}$ and after simplification one obtains
\begin{eqnarray}
\langle \alpha^{2}r^{2}\rangle &=& \sqrt{\frac{n'_{r}!n_{r}!}{{(n_{r}+|l|)!(n'_{r}+|l|)!}}} \int_{0}^{\infty}\!\!\,\mbox{d}(\alpha^{2}_{}r^{2}_{}) (r^{2}_{}\alpha^{2}_{})^{|l|+1}\nonumber\\
&\times&e^{-r_{}^{2}\alpha_{}^{2}}
L^{|l|}_{n_{r}}(r_{}^{2}\alpha_{}^{2})L^{|l|}_{n'_{r}}(r_{}^{2}\alpha_{}^{2}).\nonumber
\end{eqnarray}
Using the orthogonality relation for Associated Laguerre polynomials, one finally obtains
\begin{eqnarray}
\langle \alpha^{2}r^{2}\rangle&=& \sqrt{\frac{n'_{r}!(n_{r}+|l|)!}{{n_{r}!(n'_{r}+|l|)!}}}(2n_{r}+|l|+1).
\end{eqnarray}
For $l=0$, the right hand side of the above equation reduces to ($2n_{r}+1$).
\bibliography{TwoParticles}
\end{document}